 \documentclass[twocol]{ametsocV6.1}

\usepackage{enumitem}
\usepackage{makecell}
\usepackage[normalem]{ulem}
\usepackage{amsmath}
\usepackage{makecell}
\usepackage{nicefrac}
\usepackage{cuted,mathtools}
\usepackage{hyperref}

\newcommand{\blave}[1]{\langle #1 \rangle^1}

\newcommand{\iave}[1]{\langle #1 \rangle^i}

\usepackage{hyperref}
\hypersetup{
    colorlinks=true,
    linkcolor=blue,
    urlcolor=blue,
    linktoc=all,
    citecolor=blue
           }

\title{Top- and bottom-heavy vertical velocity structures: physical modes of layered atmospheric models}

\authors{Fiaz Ahmed,\aff{a}\correspondingauthor{Fiaz Ahmed, fiaz@ucla.edu} 
J. David Neelin,\aff{a} 
}

\affiliation{\aff{a}{Department of Atmospheric and Oceanic Sciences, University of California Los Angeles}\\}

\abstract{Tropical East and West Pacific Oceans display differences in their vertical velocity (or omega) profiles. The East Pacific is characterized by bottom-heavy profiles, while the West Pacific is characterized by top-heavy profiles. Although inter-basin differences in the horizontal SST gradient are known to be important, physical reasons for why these omega structure variants exist are not fully understood. This question is addressed using a steady, linear model on an $f$-plane with $n$ atmospheric layers. Convection and radiation are parameterized as linear responses to thermodynamic perturbations with convective nonlinearity approximated by  convection on/off regimes. The free (or eigen) modes of the model yield vertical structures resembling the observed baroclinic modes of the tropical atmosphere, with each mode associated with a characteristic horizontal scale (the eigenvalue). In the standard parameter regime, the first-baroclinic mode has a large spatial scale ($\sim$ 1500 km) while the second-baroclinic mode has a smaller spatial scale ($\sim$ 250 km). When the model is forced with a strong- and weak-gradient surface temperature ($T_s$) patterns, the resulting omega profiles assume bottom- and top-heavy structures respectively---mimicking the observed differences between East and West Pacific Oceans. Additional dependence on the magnitude of the Coriolis force is also observed. The connection between the vertical structure and the horizontal scale of the baroclinic modes explains why a strong-gradient $T_s$ profile projects strongly onto the second-baroclinic mode yielding bottom-heavy omega profiles in the eastern Pacific, while a weak-gradient $T_s$ profile projects strongly onto the first-baroclinic mode, yielding top-heavy omega profiles typical of the Western Pacific.}
\newcommand{\figref}[1]{Fig. \ref{#1}}
\newcommand{\figsref}[1]{Figs. \ref{#1}}

\begin{document}

%% Necessary!
\maketitle

%%%%%%%%%%%%%%%%%%%%%%%%%%%%%%%%%%%%%%%%%%%%%%%%%%%%%%%%%%%%%%%%%%%%%
% SIGNIFICANCE STATEMENT/CAPSULE SUMMARY
%%%%%%%%%%%%%%%%%%%%%%%%%%%%%%%%%%%%%%%%%%%%%%%%%%%%%%%%%%%%%%%%%%%%%
%
% If you are including an optional significance statement for a journal article or a required capsule summary for BAMS 
% (see www.ametsoc.org/ams/index.cfm/publications/authors/journal-and-bams-authors/formatting-and-manuscript-components for details), 
% please apply the necessary command as shown below:
%
% Significance Statement (all journals except BAMS)
%
\statement
%
%% Capsule (BAMS only)
%%
%\capsule
%       Enter BAMS capsule here, no more than 30 words. See \url{www.ametsoc.org/index.cfm/ams/publications/author-information/formatting-and-manuscript-components/#capsule} for details.
% 
%% * * If using twocol mode, you will need to use the commands "twocolsig" and "twocolcapsule" in place of "sig" and "capsule"
%%      to ensure that the text box correctly spans across both columns.

%%%%%%%%%%%%%%%%%%%%%%%%%%%%%%%%%%%%%%%%%%%%%%%%%%%%%%%%%%%%%%%%%%%%%
% MAIN BODY OF PAPER
%%%%%%%%%%%%%%%%%%%%%%%%%%%%%%%%%%%%%%%%%%%%%%%%%%%%%%%%%%%%%%%%%%%%%
%
\section{Introduction}

Typically, precipitating regions display upward motion through much of the atmospheric column. However, the vertical structure of this upward motion can vary within the tropics. These variations are most starkly illustrated when comparing vertical velocity (or omega) profiles between the tropical East and West Pacific Oceans. During precipitating times, the composite omega profiles in the East Pacific exhibit a maximum above the boundary layer, while those in the West Pacific exhibit a maximum in the upper troposphere \citep{back2006geographic,back2009simple,fuchs2020otrec2019, huaman2022assessing, bernardez2024integrating}. These canonical East and West Pacific profiles are termed `bottom-heavy' and `top-heavy' profiles respectively. These variants have been independently documented using data from reanalyses,  soundings and satellite retrievals, although there exist quantitative variations among these products \citep{huaman2018assessing}. Closely related to these top- and bottom-heavy structures are the two leading empirical orthogonal functions (EOFs) of omega profiles, which account for a bulk of the variance in tropical omega profiles \citep{hagos2010estimates,handlos2014estimating,inoue2020vertical}.

The vertical structure of omega controls the gross moist stability \citep{neelin1987modeling, raymond2009mechanics,inoue2015gross}---a parameter that measures the efficiency with which convection exports column moist static energy. The gross moist stability impacts both mean and transient climate phenomena. For instance, theories for convective life-cycle evolution \citep{inoue2017gross,maithel2022moisture}, the Madden Julian Oscillation \citep{sobel2013moisture, adames2016mjo} and the ITCZ width \citep{ahmed2023process} all hinge on the smallness of the gross moist stability relative to the dry stability. Even small inter-model variations in the gross moist stability generate a large inter-model spread in the ITCZ width \citep{ahmed2023process}.

The shape of the omega profile is also tightly linked to makeup of the cloud population---specifically, the fraction of congestus, deep convective and stratiform clouds \citep{johnson1999trimodal}. Deep convective clouds display a first-baroclinic latent heating profile (assumed equivalent to the omega profile) with a single mid-level tropospheric maximum \citep{johnson1984partitioning}. Congestus clouds, on the other hand, display a second-baroclinic latent heating profile  with low-level heating and upper-level cooling \citep{schumacher2004tropical, takayabu2010shallow}. Stratiform clouds also display a second-baroclinic heating profile, but with low-level cooling and upper-level heating \citep{mapes1993gregarious,mapes1995diabatic}. These baroclinic modes can be used as basis functions to construct omega profiles for a given cloud population \citep{schumacher2004tropical, jakob2008precipitation, khouider2006simple}. However, most climate models can simulate the gross differences between the omega profiles in the East and West Pacific \citep{back2006geographic,chen2016impacts,annamalai2020enso}, despite differences in their treatment of clouds. This suggests that physical mechanisms independent of the cloud population likely govern the top- versus bottom-heaviness of omega profiles in the tropics. 

Our physical understanding for why omega profiles assume a top- or bottom-heavy structure remains incomplete. Empirical evidence, however, suggests a role for the two dominant pathways by which tropical convection is forced \citep{back2009simple}. When horizontal sea surface temperature (SST) gradients force strong near-surface convergence \citep{lindzen1987role, stevens2002entrainment, raymond2006dynamics}---as in the tropical East Pacific---the resulting convection is predominantly bottom-heavy. Over regions with weak SST gradients but warm waters---such as the tropical West Pacific---the resulting convection is thought to respond to atmospheric instability perturbations, and assume a top-heavy profile. A statistical model linking top-heaviness to instability, and bottom-heaviness to surface convergence can reproduce the precipitation climatology in current and future climates \citep{back2009simple, duffy2020importance}. Detailed mechanistic details for why these empirical relationships exist are unavailable---despite it being known that models that assume boundary layer temperature dominance and models that deep-convectively adjust temperature through the troposphere  can be cast in comparable form, each driven by SST patterns \cite[e.g.,][]{Neelin1989a,YuNeelin1997}. In addition to SST gradients, the lower-tropospheric static stability \citep{bernardez2024integrating} has also been shown to govern the bottom-heaviness of convection \citep{herman2014wtg,sessions2015convective}.

This study aims to elaborate on the mechanistic details of how top- and bottom-heavy omega profiles emerge, using a simple, linear model of tropical dynamics. Section 2 elaborates on the model setup. Section 3 examines the free modes of the linear model and shows that they correspond to observed baroclinic modes. Section 5 examines the forced solutions of the mode. Section 5 presents a full solution to the model forced by a prescribed surface temperature forcing. The horizontal scale of the forcing is shown to determine the top-heaviness of the omega solution. Section 6 ends with a summary and related discussion.

% \subsection{Key Points}

% -Moist static energy budgets to show how the bottom-heavy convection is maintained against amplification.

\section{Model Setup}\label{sec:model_setup}

\subsection{Overview}
In this section, we present the governing equations for an $n$-layered model forced by a surface temperature forcing. The equations are finite-dimensional and linear in the unknown state variables. They can therefore be represented in matrix form:
\begin{align}\label{eq:matrix_form_overview}
    \boldsymbol{A}\boldsymbol{x} - \nabla^2\boldsymbol{x} = \boldsymbol{f_s},
\end{align}
where $\boldsymbol{A}$ is a parameter matrix with dimension $2n$, $x$ is a the unknown state vector and $\boldsymbol{f_s}$ is the forcing. The horizontal Laplacian operator is $\nabla^2$

The first key result is that the eigenmodes of the parameter matrix $\boldsymbol{A}$ resemble the observed baroclinic modes of the tropical atmosphere. The second key result is the scale of the forcing---which enters the problem through the $\nabla^2$ term---controls the properties of the forced solution. A large-scale (weak gradient) forcing tends to project more strongly onto the first-baroclinic mode and produce top-heavy omega solutions. A small-scale (strong-gradient) forcing tends to produce more bottom-heavy solutions. The rest of this Section discusses the specifics of the derivation that results in \eqref{eq:matrix_form_overview}.

\subsection{Architecture}

\begin{figure}[h]
\centerline{\includegraphics[height=9.5 cm]{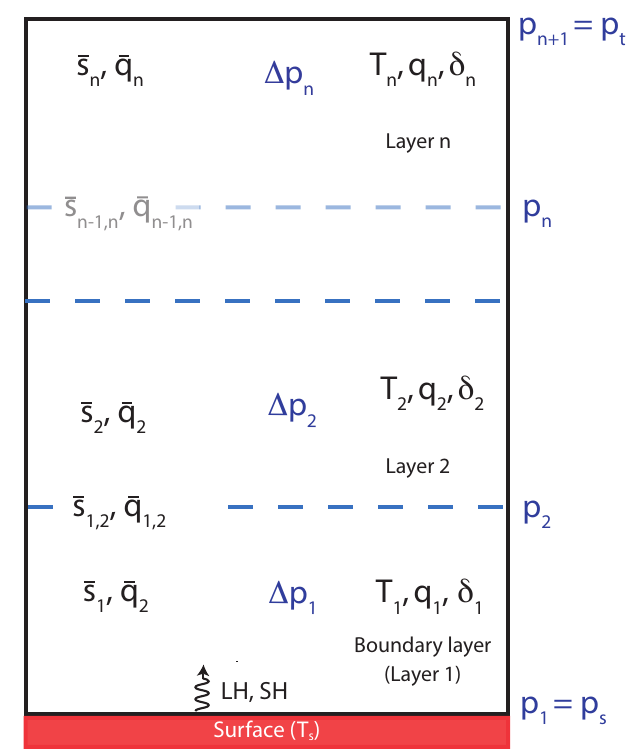}}
 \caption{The linear model with $n$ vertical layers. For each layer, the spatially-uniform background quantities are denoted by overbars, while linear perturbations around this background state are shown in black, without overbars. The pressure levels and the pressure depths for each layer are shown in blue. The surface temperature perturbation is shown in red. The surface perturbation turbulent fluxes are depicted alongside the sinuous arrow.}\label{fig:model_setup}
\end{figure}

We begin with a steady model (no time variations) on an f-plane with a single horizontal dimension. Pressure is the vertical coordinate. The vertical dimension is divided into $n$ layers bounded by $n$+1 pressure levels (\figref{fig:model_setup}). Model layer $i$ is bounded by pressure levels $p_i$ and $p_{i+1}$, and has a pressure depth of $\Delta p_i = p_i - p_{i+1}$. The surface and top-of-model pressure levels are at $p_s = p_1$ and $p_t = p_{n+1}$ respectively. According to this convention, the boundary layer is contained between pressure levels $p_1$ and $p_2$, and the topmost layer levels $p_n$ and $p_{n+1}$

The unknowns in the model are linear perturbations around a spatially-uniform background state. Specifically, we will solve for linear perturbations in horizontal divergence ($\delta_i$), temperature ($T_i$) and specific humidity ($q_i$) within each layer. The horizontal divergence $\delta_i$ is constant layer $i$, but the temperature and specific humidity perturbations vary with fixed vertical structures $a_i(p)$ and $b_i(p)$ respectively. These vertical structure functions $a_i(p)$ and $b_i(p)$ both equal unity at the base of layer $i$. That is, $a_i(p_i) = b_i(p_i) = 1$. The model is externally forced by a prescribed surface temperature perturbation $T_s$.

The variables characterizing the spatially-uniform background state are identified using overbars. The background temperature and specific humidity values are $\bar{T}(p)$ and $\bar{q}(p)$. The background, layer-averaged dry static energy (dse) and specific humidity within layer $i$ are denoted by $\bar{s}_i$ and $\bar{q}_i$ respectively. The subscript $i$ indicates the layer over which the vertical averages are taken. The values of background dse and specific humidity on the pressure level $p_i$ are given by $\bar{s}_{i-1,i}$ and $\bar{q}_{i-1,i}$. These interfacial variables prove important to track when considering vertical transport between layers. The background divergence---and therefore omega---is uniformly zero everywhere. The total (background plus perturbation) divergence, temperature and specific humidity values within any layer $i$ are reconstructed using:
\begin{align}
    \delta (p) = \delta_i\label{eq:delta_defn}\\
    T(p) = a_i(p)T_i + \bar{T}(p) \label{eq:T_vert_struct}\\
    q(p) = b_i(p)q_i + \bar{q}(p) \label{eq:q_vert_struct},
\end{align}
where $p \in [p_i, p_{i+1}]$. 

The specifics of this model are inspired by the Quasi-equilibrium Tropical Circulation Model \citep[QTCM;][]{neelin2000quasi,sobel2006boundary}. A major difference exists in the treatment of the vertical dimension. In contrast to assuming prefixed vertical modes \cite[e.g., ][]{neelin2000quasi,mapes2000convective}, we consider $n$ independent pressure levels in the vertical \citep[e.g., following][]{wang1993simple}. This allows solutions resembling the observed top- and bottom-heavy modes to emerge from the model than being built into it.

\subsection{General governing equations}

Given $n$ vertical layers, the model contains $3n$ unknowns: $\delta_i$, $T_i$ and $q_i$ within each layer. To solve the linear system, we require $3n$ linear equations. These equations naturally arise from the linearized horizontal momentum, energy and specific humidity equations within each layer. We now provide a brief derivation of these equations, with several additional details outlined in the Supplement.

\subsubsection{Momentum equations}

We assume that the vertical velocity $\omega$ is zero at the surface, that is, $\omega\rvert_{p_1}=0$. 
This constraint, along with mass continuity provides an expression for the vertical velocity $\omega$ on the pressure level $p_i$, where $i>1$:
\begin{align}\label{eq:omega_i+1_defn}
	\omega\rvert_{p_i} = \sum\limits^{i}_{j=1}\delta_j \Delta p_j,
\end{align}
where the constant $\delta_i$ condition within layer $i$ is used. We further make the rigid lid assumption such that the vertical velocity vanishes at the model top, that is, $\omega\rvert_{p_{n+1}} = 0$. Using \eqref{eq:omega_i+1_defn}, this condition is:
\begin{align}\label{eq:rigid_lid}
    \sum\limits^{n}_{j=1}\delta_j \Delta p_j = 0.
\end{align}
Using \eqref{eq:omega_i+1_defn} and \eqref{eq:rigid_lid} allows us to write the governing equations for horizontal divergence within each layer:
\begin{align}
	\delta_1 = \tau_1\nabla^2\left(\frac{c^2_1(1-w_{1})}{\bar{s}_1}T_1+ \sum\limits^{n}_{i=2}\frac{c^2_iw_i}{\bar{s}_i} T_i\right) \label{eq:bl_div}\\
    \delta_i = -\tau_i\nabla^2\left(\frac{c^2_1w_1}{\bar{s}_1}T_1 + \frac{c^2_i}{\bar{s}_i}T_{i}  - \sum\limits^{n}_{\substack{j=2\\ j\ne i}}\frac{c^2_jw_j}{\bar{s}_j}T_{j}\right),\ \mathrm{\ i>1}\label{eq:ft_div}.
\end{align}
The parameters $c_i$ and $\tau_i$ are the phase speed and timescale parameters that are functions of the background state. The parameter $w_{i}$ is a vertical weighting parameter. The definitions for these these parameters is provided in Table \ref{tab:background_parameters} and discussed further in the Supplement. 

The defining feature of \eqref{eq:bl_div} and \eqref{eq:ft_div} is the strong coupling between layers. The horizontal divergence in layer $i$ depends on the horizontal Laplacian $\nabla^2$ of temperature perturbations within every layer of the column. This coupling arises for two reasons. Firstly, the geopotential perturbation within any layer $i$ depends on the temperature of all layers underneath. Secondly, the rigid lid assumption \eqref{eq:rigid_lid} imposes a constraint on the column-integrated horizontal divergence. A physical interpretation of this latter condition is that the vertical wave energy reflected off the rigid lid \citep{chumakova2013leaky, edman2017beyond} transmits information from the upper to lower layers.

To properly capture SST-induced boundary layer convergence, the  geopotential perturbation at the boundary layer top must be included. This term can be indirectly parameterized using an adjustment process \citep[e.g., the `back-pressure' term in][]{lindzen1987role} or prescribed as an external boundary term \citep{back2009relationship}. The expression \eqref{eq:bl_div} shows that to completely account for this term, we must capture the dynamics of the entire troposphere. It is also worth noting that the boundary layer convergence in \eqref{eq:bl_div} is not directly forced by the SST. It is instead driven by tropospheric temperature perturbations, including those in the boundary layer \citep{sobel2006boundary,gonzalez2024dynamical}. 

The parameters $\tau_i$, $w_i$ and $c_i$ are all positive (see Supplement). Therefore  \eqref{eq:bl_div} implies that perturbation warming in both the boundary layer and the free troposphere will contribute to boundary layer convergence. In any free-tropospheric layer, \eqref{eq:ft_div} implies that both local warming (layer $i$) and boundary layer warming (layer $i=1$) support divergence. However, non-local free-tropospheric warming---the last right-hand side term in \eqref{eq:ft_div}---tends to induce convergence. From \eqref{eq:bl_div} and \eqref{eq:ft_div}, we deduce that boundary layer warming tends to induce bottom-heavy omega profiles by supporting convergence within the boundary layer, and divergence within all free-tropospheric layers. However, strong upper-tropospheric warming can shifting the divergence layers aloft and support a more top-heavy profile. In other words, the top- and bottom-heaviness of omega profiles in this model is dictated by the vertical profile of temperature perturbations. However, this effect is strongly modulated by the vertical weighting parameters $w_{i}$ (Table \ref{tab:background_parameters}), which have latitudinal dependence. Specifically, $w_{1}$ increases with latitude, while $w_{i}$ decreases with latitude for $i>1$. The impact of these changes is that the same temperature profile will induce a more bottom-heavy omega profile off the equator. 

% \begingroup
\renewcommand{\arraystretch}{3.0}
\begin{table*}[t]  %FA: make table in single-column
\small
\caption{Background State Parameters. Each parameter is either defined numerically (for physical constants) or symbolically, as applicable. }\label{tab:background_parameters}
\begin{center}
\begin{tabular}{ccccc}
\topline
{Parameter} & {Description} &  \makecell{Definition} & Units  & Remarks \\
\midline
    $(R_d, R_v)$ & \makecell{Gas constant of\\ (dry air, water vapor)} & \makecell{(287, 461)} & $\mathrm{J K^{-1} kg^{-1}}$ &-  \\
    $L_v$ & \makecell{Latent heat of\\ vaporization for water } & \makecell{$2.4\times10^6$} & $\mathrm{Jkg^{-1}}$ &-  \\
    $c_{pd}$ & \makecell{Heat capacity of dry air } & \makecell{$1004$} & $\mathrm{Jkg^{-1}K^{-1}}$ &-  \\
    $a^+_i(p)$ & \makecell{Geopotential\\ vertical structure} & \makecell{ $\int^{p}_{p_{i+1}} a_i[1+ (R_d/R_v) \bar{q}]\ d\ln p'$}  & - & \makecell{Function of $a_i(p)$ and\\ background humidity $\bar{q}_i(p)$.} \\
    $c_i$ &  \makecell{Phase speed\\within layer $i$} & \makecell{$\sqrt{R_d\iave{a^+_i}\bar{s}_i}$} & $\mathrm{ms^{-1}}$  &\makecell{$\iave{}$ is the vertical averaging\\
    operation within layer $i$.} \\ 
    $\epsilon_i$ &\makecell{Rayleigh friction\\
    strength in layer $i$} & - & $\mathrm{s^{-1}}$& \makecell{Linear drag parameterizes\\ friction \citep{deser1993diagnosis, wu2000rayleigh}}. \\
    $f$ & Coriolis parameter & $2\Omega \sin \theta_0$ & $\mathrm{s}^{-1}$ & \makecell{The latitude is $\theta_0$. Earth's rotation\\ rate $\Omega = 7.2921\times10^{-5}\mathrm{rad\ s^{-1}}$.}\\
    $\tau_i$ & \makecell{Rotational-frictional\\timescale in layer $i$} & $\epsilon_i(\epsilon^2_i + f^2)^{-1}$ & s &\makecell{Rotation enters the system\\
    only through $\tau_i$.}\\
    $w_{i}$ & \makecell{Rotational-frictional\\ vertical weighting} & $\tau_i\Delta p_i\left(\sum\limits^{n}_{j=1}\tau_j\Delta p_j\right)^{-1}$ & - &\makecell{A positive fraction
    whose\\ sum across layers equals 1.}\\    
    $\kappa_s$ &  \makecell{Surface flux timescale} & - & $\mathrm{s^{-1}}$ & \makecell{Controls the strength of surface \\disequilibrium adjustment}  \\ 
    $\gamma_s$ & \makecell{Temperature to saturation\\ specific humidity coefficient}  & $L_v\bar{q}_s(R_v\bar{T}_s^2)^{-1}$ & - & \makecell{$\bar{T}_s$ and $\bar{q}_s$ are background surface\\ temperature (K) and saturation specific\\ humidity (K) respectively.}  \\ 
    $\epsilon_\mathrm{mix}$ &  \makecell{Turbulent\\mixing timescale}  & - &  $\mathrm{s^{-1}}$ & \makecell{Vertical mixing at layer interfaces.\\ Following \cite{sobel2006boundary}.} \\
    $(\nu_T, \nu_q)$ &  \makecell{(Temperature, moisture)\\ diffusion coefficient} & 
    - &$\mathrm{m^2s^{-1}}$ & \makecell{Non-linear  flux convergence \\ \citep{sobel2006boundary}.}  \\
    $f_p$ &  \makecell{Fraction of precipitating times\\ in the convective zone} &  0.2 & - & \makecell{Convective adjustment timescales \\reduce upon time averaging\\ \citep{ahmed2020deep}} \\
    $r_c$ &  \makecell{Cloud-radiative\\heating factor}  & 0.2 &  - & \makecell{Cloud-radiative heating is assumed \\ proportional to convective heating\\\citep{su2002teleconnection,kim2015role}.} \\
 \botline 
 \end{tabular} 
\end{center}
\end{table*} %FA: make table in single-column
% \endgroup
% \clearpage
\subsubsection{Thermodynamic equations}

We now present the governing equations for the horizontal components of temperature $T_i$ and specific humidity $q_i$ within layer $i$. Recall that the corresponding vertical structures are given by $a_i(p)$ and $b_i(p)$ from \eqref{eq:T_vert_struct} and \eqref{eq:q_vert_struct} respectively. In the subsequent expressions, both $T_i$ and $q_i$ have units of $K$, after scaling the specific humidity equation by $L_v/c_{pd}$. The layer-averaged dry static energy and specific humidity equations for the boundary layer ($i=1$) are:
 \begin{strip}  %FA: turn off in single-column mode
\begin{align}
     \underbrace{-\delta_1(\bar{s}_{1,2} - \bar{s}_1)}_{\substack{\text{net adiabatic}\\\text{cooling}}}  = \underbrace{\kappa_s (T_s - T_1)}_{\substack{\text{surface sensible }\\ \text{heat flux}}}+\underbrace{h_1}_{\substack{\mathrm{convective}\\ \mathrm{heating}}} + \underbrace{r_1}_{\substack{\mathrm{radiative}\\ \mathrm{heating}}} + \underbrace{\nu_T\blave{a_1}\nabla^2T_{1}}_{\substack{\text{temperature}\\\text{diffusion}}}+\underbrace{\epsilon_\mathrm{mix}[T_2 - a_1(p_2)T_1]}_\text{turbulent mixing}\label{eq:bl_dse_eqn}\\
	 \underbrace{-\delta_0(\bar{q}_{1,2} - \bar{q}_1)}_{\substack{\text{net vertical}\\\mathrm{moistening}}}   = \underbrace{\kappa_s (\gamma_sT_s - q_1)}_{\substack{\text{surface latent }\\ \text{heat flux}}}+\underbrace{d_1}_{\substack{\mathrm{convective}\\ \mathrm{drying}}} +
\underbrace{\nu_q\blave{b_1}\nabla^2q_{1}}_{\substack{\text{moisture}\\\text{diffusion}}}+\underbrace{\epsilon_\mathrm{mix}[q_2 - b_1(p_2)q_1]}_\text{turbulent mixing}.\label{eq:bl_vap_eqn}
\end{align}
 \end{strip}  %FA: turn off in single-column mode
 
The derivations for \eqref{eq:bl_dse_eqn} and \eqref{eq:bl_vap_eqn} are shown in the Supplement, with a brief interpretation provided here. In \eqref{eq:bl_dse_eqn}, the boundary layer convergence $\delta_1$ acts against a background static stability to produce a net adibatic cooling. The relevant static stability measure is
$\bar{s}_{1,2} - \bar{s}_1$, which accounts for both adiabatic cooling within and the dse flux out of the boundary layer. This net adiabatic cooling is balanced by a sum of surface sensible flux,  diabatic heating that includes convective and radiative heating \citep{yanai1973determination}, horizontal temperature diffusion and linear turbulent mixing due to entrainment at the boundary layer top. The relevant static stability for this dse balance is $\bar{s}_{0,1} - \bar{s}_0$, which is the difference between the boundary layer top dse and the boundary layer averaged dse. The boundary layer averaged specific humidity balance \eqref{eq:bl_vap_eqn} is analogous to \eqref{eq:bl_dse_eqn}---with convective drying replacing diabatic heating. 

\clearpage

A similar pair of thermodynamic equations are derived for free-tropospheric layer $i$ ($>1$):
\begin{strip}  %FA: turn off in single-column mode
\begin{align}
    \underbrace{\bar{s}_i\delta_i +\frac{\bar{s}_{i-1,i} \omega\rvert_{p_{i}}-\bar{s}_{i,i+1}\omega\rvert_{p_{i+1}}}{\Delta p_i}}_{\substack{\text{net adiabatic cooling}}} 
= \underbrace{h_i}_{\substack{\mathrm{convective}\\ \mathrm{heating}}} + \underbrace{r_i}_{\substack{\mathrm{radiative}\\ \mathrm{heating}}} + \underbrace{\nu_T\iave{a_i}\nabla^2T_i}_{\substack{\text{temperature}\\\text{diffusion}}} +\underbrace{\frac{\Delta p_1}{\Delta p_i}\epsilon_\mathrm{mix}\left[a_{i-1}(p_i)T_{i-1} -(1 + a_{i}(p_{i+1}))T_{i}  + T_{i+1} \right]}_{\substack{\text{turbulent mixing}}}\label{eq:ft_dse_eqn}
\end{align}
\begin{align}
\underbrace{\bar{q}_i\delta_i +\frac{\bar{q}_{i-1,i} \omega\rvert_{p_{i}}-\bar{q}_{i,i+1}\omega\rvert_{p_{i+1}}}{\Delta p_i}}_{\substack{\text{net vertical moistening}}}  
= \underbrace{d_i}_{\substack{\mathrm{convective}\\ \mathrm{drying}}} + \underbrace{\nu_q\iave{b_i}\nabla^2q_i}_{\substack{\mathrm{moisture}\\\mathrm{diffusion}}}
+ \underbrace{\frac{\Delta p_1}{\Delta p_i}\epsilon_\mathrm{mix}\left[b_{i-1}(p_i)q_{i-1} -(1 + b_{i}(p_{i+1}))q_{i}  + q_{i+1} \right]}_{\substack{\text{turbulent mixing}}}\label{eq:ft_vap_eqn}.
\end{align}
\end{strip}

In \eqref{eq:ft_dse_eqn}, the net adiabatic cooling term on the left-hand side includes both within-layer (the leftmost term) and interfacial (the second-from-left term) contributions. The turbulent mixing term on the right-hand side accounts for mixing across the layer top and bottom boundaries. The boundary layer equations \eqref{eq:bl_dse_eqn} and \eqref{eq:bl_vap_eqn} can be recovered from \eqref{eq:ft_dse_eqn} and \eqref{eq:ft_vap_eqn} respectively. This is achieved by setting $i=1$, using the constraint $\omega\rvert_{p_1}=0$ and replacing near-surface linear turbulent mixing with bulk surface flux parameterizations.

The surface temperature perturbation $T_s$ enters the surface flux terms in \eqref{eq:bl_dse_eqn} and \eqref{eq:bl_vap_eqn}, and the radiative heating term in \eqref{eq:bl_dse_eqn} and \eqref{eq:ft_dse_eqn}.

\subsubsection{Convective and radiative closures}

To close the thermodynamic equations, we specify closures for the convective and radiative heating terms in \eqref{eq:bl_dse_eqn} and \eqref{eq:ft_dse_eqn}, and the convective drying term in \eqref{eq:bl_vap_eqn} and \eqref{eq:ft_vap_eqn}. To retain linearity, we seek linear closures for convection and radiation in terms of the thermodynamic perturbations $\{T_i\}$ and $\{q_i\}$. 

We first define column vectors of perturbation  temperature $\boldsymbol{x_T}$ and water vapor $\boldsymbol{x_q}$ for an $n$-layered system:
\begin{align}
\boldsymbol{x_T} = \begin{bmatrix}
    T_1 & T_2 & \dots & T_n
\end{bmatrix}^T \label{eq:xT_defn}\\
\boldsymbol{x_q} = \begin{bmatrix}
    q_1 & q_2 & \dots & q_n
\end{bmatrix}^T, \label{eq:xq_defn}
\end{align}
and column vectors of perturbation convective heating $\boldsymbol{h}$, radiative heating $\boldsymbol{r}$ and convective drying $\boldsymbol{d}$:
\begin{align*}
\boldsymbol{h} = \begin{bmatrix}
    h_1 & h_2 & \dots & h_n
\end{bmatrix}^T\\
\boldsymbol{r} = \begin{bmatrix}
    r_1 & r_2 & \dots & r_n
\end{bmatrix}^T\\
 \boldsymbol{d} = \begin{bmatrix}
    d_1 & d_2 & \dots & d_n
\end{bmatrix}^T.
\end{align*} 
Linear closures for the diabatic and convective drying terms can now be expressed in block-matrix form:
\begin{align}
    \boldsymbol{h} = \begin{bmatrix}
        \boldsymbol{H_T} & \boldsymbol{H_q}
    \end{bmatrix}\begin{bmatrix}
        \boldsymbol{x_T} \\ \boldsymbol{x_q}
    \end{bmatrix}\label{eq:conv_heating}\\
    \boldsymbol{r} = \begin{bmatrix}
        \boldsymbol{R_T} & \boldsymbol{R_q}
    \end{bmatrix}\begin{bmatrix}
        \boldsymbol{x_T} \\ \boldsymbol{x_q}
    \end{bmatrix} + T_s\boldsymbol{r_s}\label{eq:rad_heating}\\
    \boldsymbol{d} = \begin{bmatrix}
        \boldsymbol{D_T} & \boldsymbol{D_q}
    \end{bmatrix}\begin{bmatrix}
        \boldsymbol{x_T} \\ \boldsymbol{x_q}
\end{bmatrix}\label{eq:conv_drying}.
\end{align}
In \eqref{eq:conv_heating}, $\boldsymbol{H_T}$ and $\boldsymbol{H_q}$ are submatrices that respectively multiply $\boldsymbol{x_T}$ and $\boldsymbol{x_q}$ to yield $\boldsymbol{h}$. The submatrices $\boldsymbol{R_T}$ and  $\boldsymbol{R_q}$, and $\boldsymbol{D_T}$ and $\boldsymbol{D_q}$ similarly operate on $\boldsymbol{x_T}$ and $\boldsymbol{x_q}$ to respectively yield $\boldsymbol{r}$ and  $\boldsymbol{d}$. In \eqref{eq:rad_heating}, $r_s$ is the perturbation radiative heating vector due a unit surface temperature perturbation.
% The submatrices in \eqref{eq:conv_heating}--\eqref{eq:conv_drying} are Jacobian matrices with .

\begin{figure}[h]
\centerline{\includegraphics[width=\linewidth]{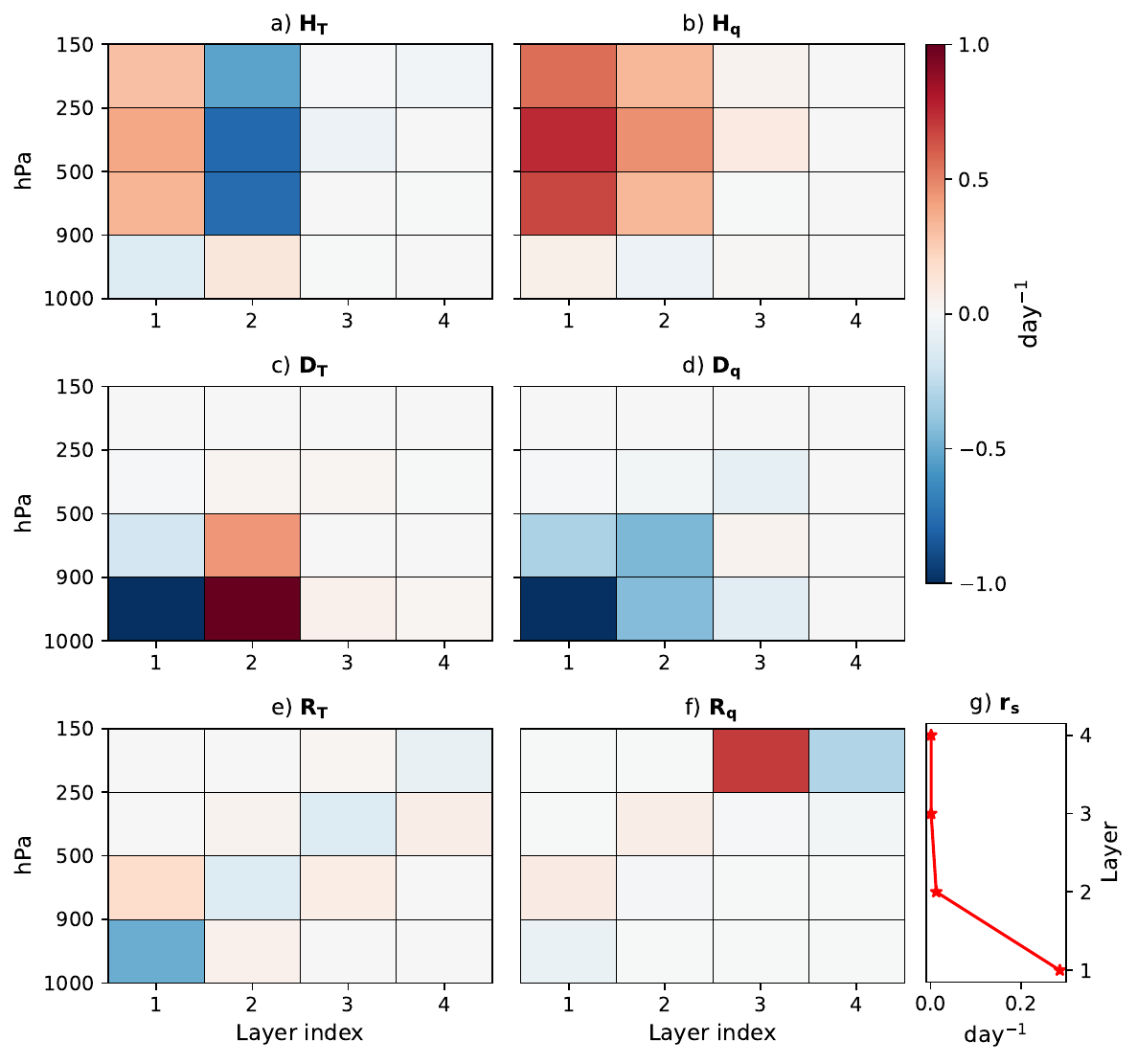}}
 \caption{The matrices of perturbation temperature and moisture contributions to convective heating (a, b), convective moistening (c, d) and clear-sky radiative heating (e, f) for the case with $n=4$ layers. Each entry in the matrix has units of inverse timescale. g) The surface temperature contribution to radiative heating. }\label{fig:radconv_matr}
\end{figure}

The steady-state linear response of a convective ensemble to thermodynamic perturbations has been documented using a cloud-resolving model \citep{kuang2010linear,kuang2018linear,kuang2024linear}. These linear response functions (LRFs) are equivalent to the convective heating and drying matrices introduced above. Numerical values for these matrices are therefore obtained from LRFs. The relationship between thermodynamic perturbations and the convective response weaken with time-averaging by a factor related to the fraction of precipitating times \citep{ahmed2020deep, nicolas2022theory}. For use within a steady-state model, the original LRF timescales were therefore scaled by a factor representing the fraction of precipitating times. This factor ($f_p$), is assumed equal to 0.2 in \figref{fig:radconv_matr}a--d. 

One instance of these LRFs is displayed in \figref{fig:radconv_matr} for $n=4$ layers, after coarse-graining the original matrices in the vertical direction. The five pressure levels that bound the four layers are shown on the y-axes of this figure. \figsref{fig:radconv_matr}a and b depict the convective heating matrices $\boldsymbol{H}_T$ and $\boldsymbol{H}_q$; \figsref{fig:radconv_matr}c and d depict the convective drying matrices $\boldsymbol{D}_T$ and $\boldsymbol{D}_q$. Each matrix entry is an adjustment timescale with units of $\mathrm{day^{-1}}$ per unit perturbation. These matrices when multiplied by $\boldsymbol{x_T}$ and $\boldsymbol{x_q}$ (with units K) yields vertical profiles of heating and drying tendencies respectively (with units of $\mathrm{K\ day^{-1}}$). Note that the matrices in \figref{fig:radconv_matr} are shown with their row ordering reversed for visual clarity, where the layer indices increase from the bottom to the top.

The main properties of LRFs are discussed in depth in \cite{kuang2010linear}, but a few key features are highlighted here. In \figref{fig:radconv_matr}a and b, we see that a boundary layer ($i=1$) warming or moistening perturbation produces convective heating in the free troposphere (\figref{fig:radconv_matr}a). Moistening or cooling in the lower-free troposphere (layer $i=1$) can similarly excite a deep heating (\figref{fig:radconv_matr}b). Thermodynamic perturbations in the upper troposphere have minimal effects on convective activity. The boundary layer experiences slight cooling when the free-troposphere is warmed, which presumably reflects the impacts of convective-scale downdrafts acting to damp boundary layer moist static energy \citep[e.g.,][]{raymond1997boundary,zuidema2017survey,schiro2018tropical}. These features are consistent with the argument that lower-tropospheric cloud buoyancy \citep{ahmed2020deep,nicolas2022theory, nicolas2024understanding} strongly modulates tropical convection. Strong convective drying can be seen accompanying these heating signatures (\figref{fig:radconv_matr}c--d), although the drying peak is strongest at the lowest levels---as observed in field data \citep{yanai1973determination,johnson2016further}.
% This separation between the heights of maximum convective heating and drying is an indication of strong vertical eddy transports . . 

The clear-sky radiative heating matrices, $\boldsymbol{R_T}$ and $\boldsymbol{R_q}$, and the surface contribution to radiative heating $\boldsymbol{r_s}|$ were obtained by perturbing Rapid Radiative Transfer Model for General Circulation Models (RRTMG), available through the CLIMLAB Python package \citep{rose2018climlab}. Specifically, the RRTMG at the reference thermodynamic state is subject to small perturbations of temperature and water vapor at every level, and the surface temperature. The difference between the perturbed and reference clear-sky radiative heating rates is divided by the size of the thermodynamic perturbation to yield $\boldsymbol{R_T}$, $\boldsymbol{R_q}$ and $\boldsymbol{r_s}$ shown in \figsref{fig:radconv_matr}e--g. Temperature variations produce a strong radiative cooling within the layer being perturbed (the cooling along diagonal in \figref{fig:radconv_matr}e), while slightly warming the adjacent layers. This differs from a simple Newtonian cooling parameterization, where the cooling is entirely local, with no impacts on adjacent layers. The clear-sky radiative response to specific humidity perturbations is strongest at the upper levels (\figref{fig:radconv_matr}f). A unit $T_s$ increase contributes most strongly to the boundary layer warming, with progressively smaller contributions to the upper layers (\figref{fig:radconv_matr}g).

The radiative heating matrices $\boldsymbol{R_T}$ and $\boldsymbol{R_q}$, like the convective matrices are also Jacobian matrices. They can also be viewed as \textit{radiative kernels} \citep{soden2008quantifying} for the reference profile. 
Cloud-radiative heating is included in the model by multiplying the convective heating $\boldsymbol{h}$ by a constant cloud-radiative factor $r_c$. This is following studies that show a linear relationship between convective and cloud-radiative heating.\citep{su2002teleconnection,kim2015role}. This treatment is admittedly simplistic, but avoids the need to develop a linearized cloud cover parameterization to obtain vertically-resolved cloud-radiative heating.

As in the momentum equations, the thermodynamic equations \eqref{eq:bl_dse_eqn}--\eqref{eq:ft_vap_eqn} are coupled in the vertical direction. Deep convection, in particular, introduce strong non-local coupling (\figref{fig:radconv_matr}a--b). Radiative heating, turbulent mixing and the dse/vapor transport across layers introduce more localized coupling across adjacent layers.

\subsection{Linear system}

We first define the column vector of thermodynamic perturbation unknowns:
\begin{align}
\boldsymbol{x} = \begin{bmatrix}
        \boldsymbol{x_T} \\
        \boldsymbol{x_q}
    \end{bmatrix}. \label{eq:x_defn}
\end{align}
To solve the combined momentum and thermodynamic equations, we substitute for $\delta_i$ from \eqref{eq:bl_div} and \eqref{eq:ft_div} into \eqref{eq:bl_dse_eqn}--\eqref{eq:ft_vap_eqn}. This eliminates the $n$ momentum equations, leaving a system with $2n$ equations. Non-linear terms in this system appear due the horizontal Laplacian $\nabla^2$, which can be grouped together. These $2n$ reduced thermodynamic equations are compactly expressed in matrix form:
\begin{align}\label{eq:matrix_form0}
    \boldsymbol{L}\boldsymbol{x} - \boldsymbol{M}\nabla^2\boldsymbol{x} = \tilde{T}_s\boldsymbol{g_s},
\end{align}
where the matrix $\boldsymbol{L}$ contains the linear terms, the matrix $\boldsymbol{M}$ contains the coefficients of the horizontal Laplacians. The term $\boldsymbol{g_s}$ is the surface forcing vector:
\begin{align}\label{eq:gs_defn}
    \boldsymbol{g_s} = \underbrace{\kappa_s\begin{bmatrix}
         \boldsymbol{e_1} \\ \gamma_s \boldsymbol{e_1}
\end{bmatrix}}_{\substack{\text{surface fluxes}}} +  \underbrace{\begin{bmatrix} \boldsymbol{r_s} \\ \boldsymbol{0} \end{bmatrix}}_{\substack{\text{surface radiation}}},
\end{align}
where $\boldsymbol{e_1}$ is the unit vector with first entry 1 and zero in the other $n-1$ entries, and $\boldsymbol{0}$ is the null vector (all zero entries). To linearize the system, we further assume that the perturbations have the form 
\begin{equation}
\boldsymbol{x} \sim \boldsymbol{\tilde{x}} \exp(kx) \ , \label{eq:expkx}
\end{equation}
where $x$ is the horizontal distance and $k$ yields a horizontal scale when real. This assumption transforms the horizontal Laplacian $\nabla^2$ into $k^2$ and allows us to write \eqref{eq:matrix_form0} as:
\begin{align}
    \boldsymbol{L}\boldsymbol{\tilde{x}} - \boldsymbol{M}k^2\boldsymbol{\tilde{x}} = \tilde{T}_s\boldsymbol{g_s},
\end{align}
which we multiply by $\boldsymbol{M}^{-1}$ on both sides to get:
\begin{align}\label{eq:matrix_form}
    \boldsymbol{A}\boldsymbol{\tilde{x}} - k^2\boldsymbol{\tilde{x}} = \tilde{T}_s\boldsymbol{f_s},
\end{align}
where $\boldsymbol{A} = \boldsymbol{M}^{-1}\boldsymbol{L}$ and $\boldsymbol{f_s} = \boldsymbol{M}^{-1}\boldsymbol{g_s}$. 

\section{Physical modes of the n-layer model}

\subsection{Baroclinic modes}

To study the free (or internal) modes of the linear system, we set the surface temperature perturbation $\tilde{T}_s=0$ in \eqref{eq:matrix_form} to get:
\begin{align}\label{eq:evp}
\boldsymbol{A}\boldsymbol{\tilde{x}} = k^2\boldsymbol{\tilde{x}},
\end{align}
which is an eigenvalue equation. Since $\boldsymbol{A}$ is a square matrix with $2n$ dimensions, we have $2n$ eigenvalues and eigenvectors (barring the case with degenerate eigenvalues). Each eigenvalue $k^2$ is associated with a spatial scale $\lambda$ given by $\lambda = 1/|\mathrm{Re}(k)|$, where Re denotes the real part. The eigenvectors that satisfy \eqref{eq:evp} contain information about temperature ($\boldsymbol{x_T}$) and water vapor ($\boldsymbol{x_q}$), from which we infer the vertical structures of horizontal divergence and omega \eqref{eq:omega_i+1_defn}, \eqref{eq:bl_div} and \eqref{eq:ft_div}. These eigenvectors, which are the internal modes of a bounded, stratified system, are baroclinic modes \citep{fulton1985vertical,yassin2022discrete}.

% The background values of dry static energy and moisture, and the vertical structure parameters in \ref{tab:parameters} were derived from a tropical sounding from the tropical West Pacific \fa{(see Supplement)}.

\begin{figure*}[h]
\centerline{\includegraphics[width=0.75\textwidth]{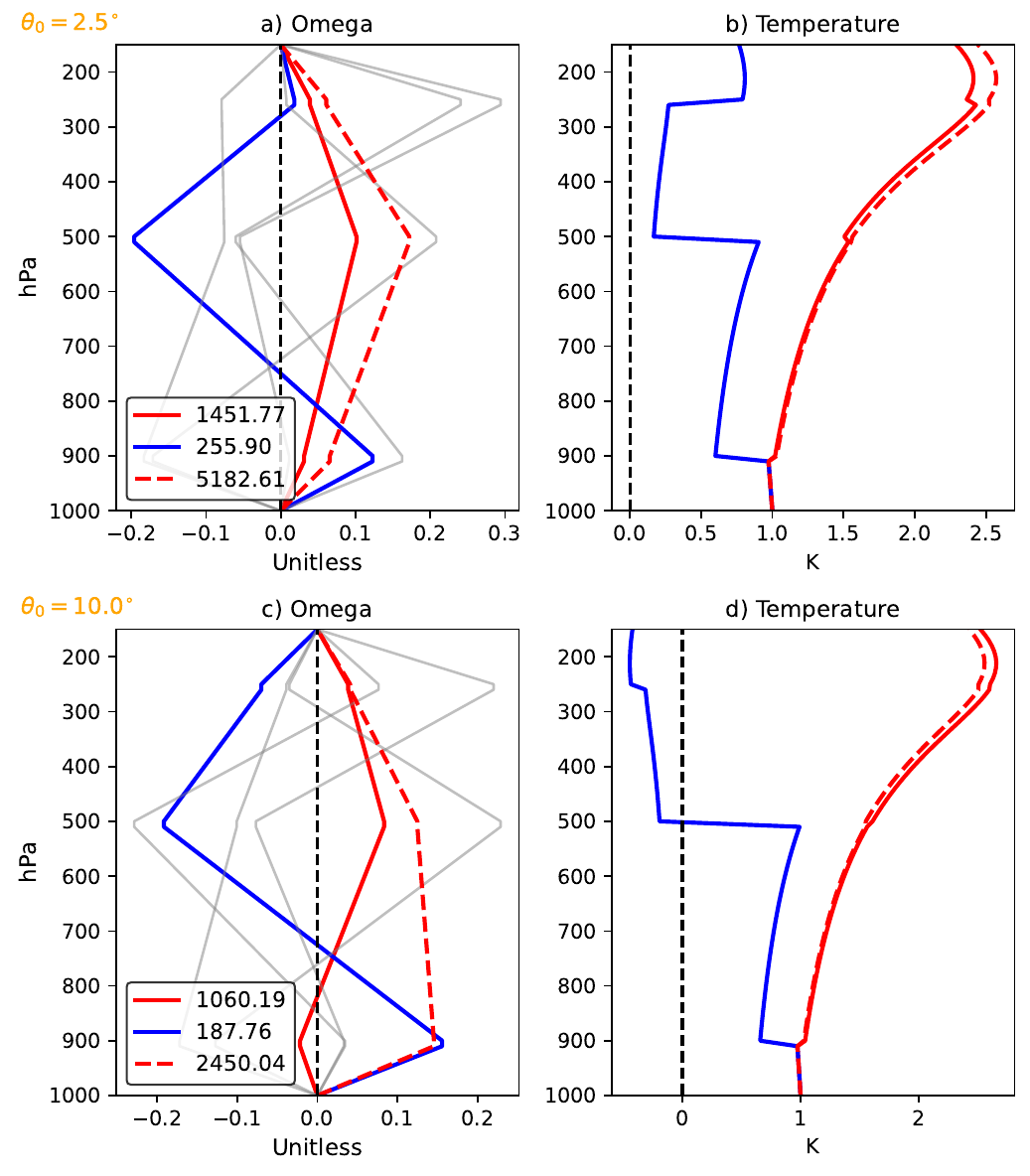}}
 \caption{The eigenmodes of the four-layer model for the (a--b) near-equatorial and (c--d) off-equatorial cases. Panels a), c) show perturbation vertical velocities normalized by the their $l^2$ norm. Panels b), d) show the corresponding temperature perturbations, normalized to have unit $T$ in boundary layer. Solid lines denote solutions with convection turned on, while dashed lines denote solutions without convection. The red and blue curves denote the two largest eigenmodes of the model; the legends in panels a and c show the eigenmode length scales in kilometers. The grey curves denote higher-order modes in the presence of convection.}\label{fig:free_modes}
\end{figure*}

% \begingroup
\renewcommand{\arraystretch}{2.0}
\begin{table}[h]
\caption{Standard parameter regime}\label{tab:parameters}
\begin{center}
\begin{tabular}{cccc}
\topline
{Var.} & {Description} &  {Value} & Units \\
\midline
    $n$ &  \makecell{Number of layers} & 4 & - \\ 
    % $\theta_0$ &  Latitude &  2.5 & degrees \\
    $\kappa_s$ &  \makecell{Surface flux timescale} & 0.32 & d$^{-1}$ \\ 
    $\epsilon_\mathrm{1}$ &  \makecell{Boundary layer\\Rayleigh friction timescale}  & 2.5 &  d$^{-1}$ \\
    $\epsilon_\mathrm{i>1}$ &  \makecell{Free-tropospheric \\Rayleigh friction timescale}  & 0.25 &  d$^{-1}$ \\
    $\epsilon_\mathrm{mix}$ &  \makecell{Turbulent vertical\\mixing timescale}  & 10 &  d$^{-1}$ \\
    $\nu_T, \nu_q$ &  \makecell{Temperature and\\ moisture diffusion} & 
    $10^6$ &$\mathrm{m^2s^{-1}}$ \\
 \botline 
 \end{tabular} 
\end{center}
\end{table}
% \endgroup

The leading eigenvectors of the linear system for a case with $n=4$ are shown in \figref{fig:free_modes}. The parameter values for various timescales and diffusion coefficients used to produce \figref{fig:free_modes} are shown in Table \ref{tab:parameters}. The system is solved both with and without convection, and for cases both near and away from the equator ($\theta_0\in\{2.5^\circ, 10^\circ\}$  latitude). To turn off convection, the convective heating and drying vectors $\boldsymbol{h}$ and $\boldsymbol{d}$ are set to zero. 

For the near-equatorial case (\figref{fig:free_modes}a), the first-baroclinic mode has a single-signed omega profile in the troposphere, and a large spatial scale ($\lambda\sim$1452 km for the case with convection). The corresponding temperature perturbations (\figref{fig:free_modes}b) are similarly deep and resemble moist adiabatic temperature perturbations. For the near-equatorial case, the second-baroclinic mode has an omega profile (\figref{fig:free_modes}a) that primarily has two signs in the troposphere (aside from a small region with another sign change at the top). The temperature perturbation component of this mode is maximum in the boundary layer (\figref{fig:free_modes}c), with a smaller free-tropospheric temperature perturbation. The spatial scale of the second-baroclinic mode is also considerably smaller ($\lambda\sim$266 km) than the first-baroclinic mode. The higher-order baroclinic modes---shown for reference in \figref{fig:free_modes}a---all have considerably smaller spatial scales than the second-baroclinic mode.  In the off-equatorial case, the first baroclinic mode is not single-signed throughout the troposphere (\figref{fig:free_modes}c)---unlike for the near-equatorial case. However, the corresponding temperature structure still resembles a moist adiabatic vertical profile (\figref{fig:free_modes}d). Compared to the near-equatorial case, the spatial scale for every baroclinic mode off the equator is smaller.

In the absence of convection, the largest free mode is single-signed throughout the troposphere. This mode appears top-heavy in $\omega$ near the equator (\figref{fig:free_modes}a) and bottom-heavy (\figref{fig:free_modes}c) off the equator, with a spatial scale that is considerably larger than its convecting counterpart. For instance the near-equatorial deep-mode length scales with and without convection are $\lambda\sim$1452 km and $\lambda\sim$5183 km respectively. 

\figsref{fig:free_modes} connects the physical modes of the steady $n$-layer model to the canonical time-dependent baroclinic modes of a tropical atmosphere with an assumed ``lid'', i.e., reflective $\omega = 0$ upper boundary condition \citep{mapes2000convective,khouider2006simple,kuang2008moisture}. Such modes are often referenced in discussing types of precipitating clouds \citep{schumacher2004tropical}. \figsref{fig:free_modes} shows that these modes are not fixed, and can vary as a function of parameters, including the latitude. The spatial scales $\lambda$ associated with each eigenvector diminishes with increasing latitude and in the presence of convection. These spatial scales are therefore interpreted as a combination of tropical Rossby radii \citep{raymond2015balanced} and damping scales. The model is sufficiently complex that there is no simple dependence of $\lambda$ on the strength of the Coriolis force or the background static stability---as for the traditional Rossby radius. Another interesting property of $\boldsymbol{A}$ is that it is asymmetric---since lower and upper tropospheric perturbations produce different responses. As a result, the resulting baroclinic modes are not orthogonal. This is in contrast to the simpler, dry systems with continuous stratification \citep[e.g.,][]{fulton1985vertical}, where the eigenmodes of the second-order ODE are constrained to be orthogonal by Sturm-Liouville theory \citep{al2008sturm}.

\section{Forced Modes}

We now seek solutions for the forced problem in \eqref{eq:matrix_form}. The eigenvectors of $\boldsymbol{A}$ form a basis in $2n$-dimensional vector space. We therefore express the forcing vector $\boldsymbol{f_s}$ and the modified state vector $\boldsymbol{\tilde{x}}$ using a linear combination of the eigenvalues $\boldsymbol{v}_i$:
\begin{align}
    \boldsymbol{f_s} = \sum_{i=1}^{2n}w_i\boldsymbol{v}_i\label{eq:fs_linear_exp}\\
    \boldsymbol{\tilde{x}} = \sum_{i=1}^{2n}\alpha_i\boldsymbol{v}_i\label{eq:x_linear_exp}
\end{align}
where the known coefficients $w_i$ and the unknown coefficients $\alpha_i$ provide the weighting for the linear combination. We assume that the forcing has the form (\ref{eq:expkx}) with decay scale given by $k_f^{-1}$. Inserting \eqref{eq:fs_linear_exp} and \eqref{eq:x_linear_exp} in \eqref{eq:matrix_form} yields a simple expression for $\alpha_i$:
\begin{align}\label{eq:alphai_coeff}
    % \sum_{i=1}^{2n}\boldsymbol{v}_i\left[\alpha_i(k^2_i-k^2) -\tilde{T}_sw_i\right] = 0\\
    \alpha_i= \tilde{T}_s\left(\frac{w_i}{k^2_i-k_f^2 }\right);\ k^2_i \ne k^2_f.
\end{align}
For a fixed forcing vector---with fixed $w_i$---the forced solution will strongly project onto eigenvector $\boldsymbol{v_i}$ if the square of the forcing scale $k^2_f$ lies close to the eigenvalue $k^2_i$. In other words, the spatial distribution of the surface temperature perturbation strongly controls the response of the $n$-layered model. 

\begin{figure*}[h]
\centerline{\includegraphics[width=0.8\textwidth]{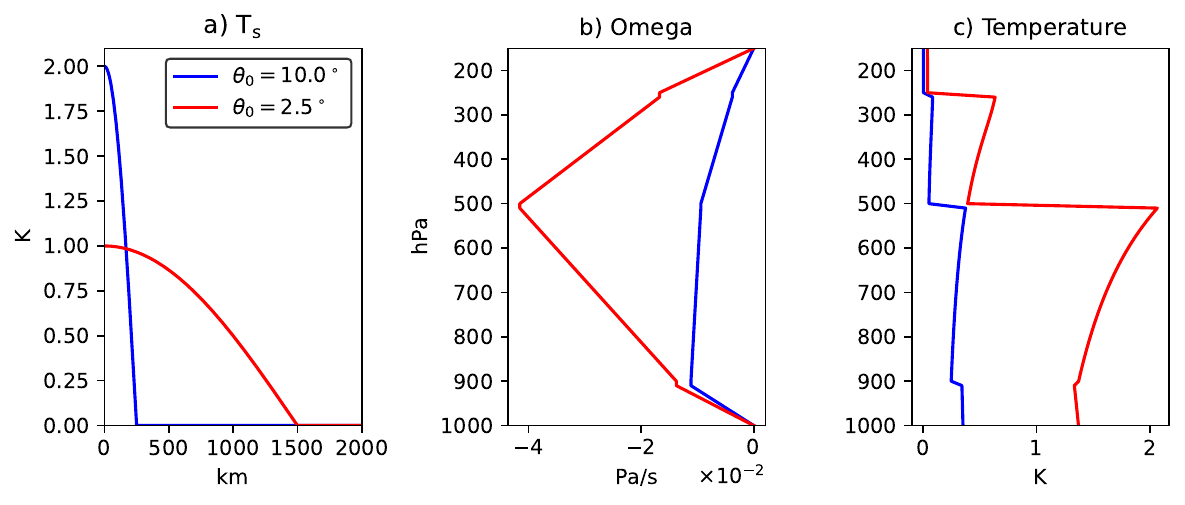}}
\caption{The surface temperature forcings (a) and the corresponding forced vertical profiles in b) omega and c) temperature.}\label{fig:forced_soln}
\end{figure*}

To illustrate this dependence on the shape of the surface temperature forcing, two different profiles are now prescribed, each with a general shape given by the cosine function:
\begin{align}\label{eq:Ts_forcing}
T_s = \begin{cases}
 T_{s0}\cos\left(\frac{\pi}{2}k_fx\right),\ x\leq x_0\\
 0,\ x>x_0.
\end{cases}    
\end{align}
where $x$ is the horizontal distance, and $T_{s0}$ is the surface temperature at $x=0$. A strong-gradient $T_s$ profile is imposed using \eqref{eq:Ts_forcing} by setting $T_{s0}=2$K, and choosing $k_f$ such that $T_s=0$ at $x_0=$250 km. A weak-gradient profile is imposed by setting $T_{s0}=1$K and allowing $T_s$ to go to zero at $x_0=$1500 km. These profiles are shown in (\figref{fig:forced_soln}a). The model is now solved for the strong-gradient $T_s$ forcing with $\theta_0=10^\circ$ (off-equator), and for the weak-gradient $T_s$ forcing with $\theta_0=2.5^\circ$ (near-equator). Perturbing both the central latitude of the $f$-plane along with the shape of the $T_s$ profile in this manner best illustrates the dichotomy between the top- and bottom-heavy profiles in the response.

The forced response in omega and temperature at $x=0$ is shown in \figsref{fig:forced_soln}b and c respectively for the particular solution  to the cosine forcing (to which matching solutions will be added below). It is apparent that the strong-gradient, off-equatorial forcing excites a bottom-heavy response in omega, while the weak-gradient, near-equatorial forcing excites a more top-heavy response. The corresponding temperature perturbations are stronger in the weak-gradient case---exceeding the peak temperature forcing in the lower troposphere---but substantially weaker in the strong gradient case. Interestingly, the upper temperature response in temperature, even in the weak-gradient case does not resemble the moist adiabatic temperature perturbations seen in \figref{fig:free_modes}. Nevertheless, it is clear that top- and bottom-heaviness in the omega profiles is strongly dependent on the shape of the imposed $T_s$ forcing---as implied by \eqref{eq:alphai_coeff}.

% The forced response will accordingly take the form $\exp\left[(i\pi/2)k_fx\right]$---provided that the forcing scale does not equal the characteristic scale of the free modes.

\section{Full solutions with a localized SST anomaly}

We now seek a full solution to a forced problem with a localized SST anomaly in a large domain with constant SST. Two variants of the surface temperature forcing are shown in \figref{fig:forced_soln}a. The full solution is a weighted sum of the free solutions and the forced response. This is represented by:
\begin{align}\label{eq:full_sol}
    \boldsymbol{x}_{\mathrm{full}} =\boldsymbol{\tilde{x}}_f\cos\left(\frac{\pi}{2}k_fx\right)+ 
\sum_{i=1}^{m}w_i\boldsymbol{\tilde{x}}_i\exp(k_ix),
\end{align}
where $\boldsymbol{\tilde{x}}_i$ is the $i^{\mathrm{th}}$ free solution (amplitude normalized),  $w_i$  is the corresponding (unknown) coefficient of linear combination, $k_i$ is the corresponding length scale. In \eqref{eq:full_sol} $m$ is the total number of free modes  included in the solution (explained below). It can be verified that $\boldsymbol{\tilde{x}}_{\mathrm{full}}$ satisfies \eqref{eq:matrix_form}. Finding the full solution now reduces to determining the weights $w_i$ (along with the particular solution amplitude $\boldsymbol{\tilde{x}}_f$).

\subsection{Convective and non-convective zones}

The $T_s$ forcing is non-zero only over a region of finite width $x_0$---as seen from \eqref{eq:Ts_forcing}. We therefore expect that convection is only active over a finite width $x_c<x_0$. The domain $x=[0,\infty)$ is thus partitioned into a convective zone with $x=[0,x_c]$ and a non-convective zone with $x=(x_c,\infty)$. In the problem setup, $x_c$ can be specified (e.g., assumed equal to $x_0$) or left as an unknown.

Within each zone, there are $2n$ free modes, each of which corresponds to a distinct value of $k_i^2$ \eqref{eq:evp}, and therefore two length scales: $\pm k_i$. Within each zone, there are a total of $4n$ possible solutions of the form $\exp(\pm k_ix)$ that appear in \eqref{eq:full_sol}, yielding a combined $8n$ unknown weights in \eqref{eq:full_sol}. We additionally impose the constraint that the solution decays as $x\to\infty$. This eliminates $2n$ modes with a positive real $k_i$ value in the non-convective zone, leaving $6n$ weights ($w_i$) to be determined. This also implies that in \eqref{eq:full_sol}, $m=4n$ in the convective zone and $m=2n$ in the non-convective zone. Note that while the solutions are found for $x>0$, symmetry about 0 implies that the convective zone can be viewed as central with decaying solutions to either side.

\subsection{Physical constraints}

To solve for the $6n$ unknowns, we require an additional $6n$ constraints, which we obtain with the following matching conditions:
\begin{enumerate}
    \item Continuity: The solution $\boldsymbol{x}_{\mathrm{full}}$ within the convective and non-convective zones must equal each other at $x=x_c$.
    \item Continuity of the first derivative: The first derivative of $\boldsymbol{x}_{\mathrm{full}}$ within the convective and non-convective zones must equal each other at $x=x_c$. 
    \item Symmetry: The convective zone solutions equal each other at $x=\pm x_c$, are thus  symmetric about $x=0$.
\end{enumerate}

The first two constraints ensure that the Laplacian of the state vector exists at $x_c$. They also guarantee that the the winds and geopotential fields at $x=x_c$ will match when approached from either side of $x_c$. Each of these above conditions provides $2n$ constraints, thus supplying a total of $6n$ constraints. 

\subsection{Strong- and weak-SST-gradient solutions}

\begin{figure*}[h]
\centerline{\includegraphics[width=\textwidth]{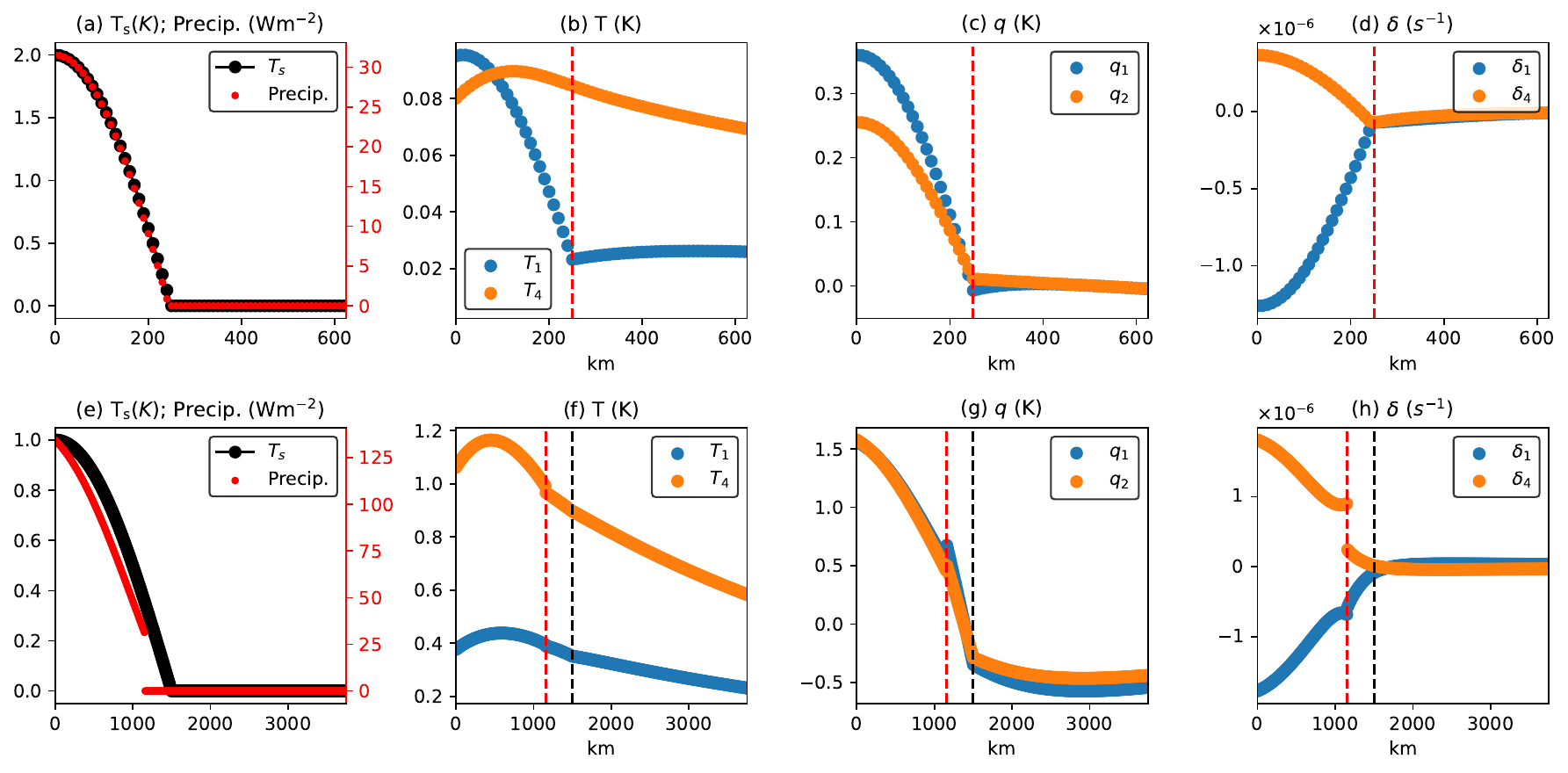}}
\caption{(a--d) Strong-gradient SST forcing, off-equatorial case: $x_0$=250 km and $\theta_0=10^\circ$ with $n=4$.
a) The surface temperature forcing (black, left $y$-axis in K) and the resulting precipitation response (red, right $y$-axis in Wm$^{-2}$). for a forcing with  b) The full solution of the temperature perturbations in the bottom ($T_1$) and upper ($T_4$) layers. c) and d) are the same as a) but for the specific humidity perturbations (in K) and horizontal divergence respectively. Panels e--h are as for a--d, but for the weak-gradient SST, near-equatorial case with  $x_0$=1500 km and $\theta_0=2.5^\circ$. The  dashed black and red vertical lines mark $x=x_0$ and $x=x_c$ respectively.}\label{fig:matching_problem}
\end{figure*}

When $x_c$ is specified, the problem is linear in the unknowns $w_i$, and can be solved such that the above constraints hold exactly. This case is illustrated in the top row of \figref{fig:matching_problem}, where the full problem is solved with a strong-gradient surface temperature forcing, with the model located off the equator. The precipitation in \figref{fig:matching_problem}a is computed by the vertical integral of the convective heating vector $\boldsymbol{h}$---which, in turn is computed using \eqref{eq:conv_heating}. When $x_c$ equals $x_0$, the convective zone with non-zero precipitation matches the forced zone, by definition. The state vector $\boldsymbol{x}_{\mathrm{full}}$ is continuous across $x=x_c$ (\figref{fig:matching_problem}b and c), and so is the horizontal divergence (\figref{fig:matching_problem}d). In \figref{fig:matching_problem}b, the boundary layer temperature perturbation ($T_1$) has a strong horizontal gradient, matching the gradient of the forcing in \figref{fig:matching_problem}a. The boundary layer solution is thus strongly tied to the forced component. $\boldsymbol{\tilde{x}}_f$. The upper-tropospheric temperature solution ($T_4$) has a shape that is considerably different from the forced profile. This is due to a strong contribution from the free modes. This upper tropospheric response continues smoothly into the non-convective zone---providing a non-local response in tropospheric temperature to a localized surface temperature forcing. The largest $w_i$ values correspond to those with the largest spatial scales (not shown), contributing to the the relatively weak temperature decay in the non-convective zone. 

In \figref{fig:matching_problem}c, the boundary layer water vapor perturbations ($q_1$) are stronger relative to the lower free troposphere ($q_2$). There exists strong boundary layer convergence with slightly smaller values of upper level divergence (\figref{fig:matching_problem}d).

The bottom row in \figref{fig:matching_problem} shows the full solution for a weak-gradient surface temperature forcing, with the model now located closer to the equator. In this case, $x_c$ is left unspecified because setting $x_c=x_0$ produced a region of strong boundary layer divergence and negative precipitation as $x\to x_0$. This spurious region (not shown) exists because no physical constraint on the sign of precipitation was initially imposed. To address this, we leave $x_c$ unspecified, which gives rise to a non-linear system of equations, since $x_c$ appears within exponentials $\sim \exp(kx_c)$. We then include an additional physical constraint of  zero precipitation at $x=x_c$. Unlike the case with specified $x_c$, this system does not have a closed-form solution. The system is therefore solved with an iterative, non-linear solver to produce joint estimates for both $w_i$ and $x_c$ (\figsref{fig:matching_problem}e--h). The red dashed lines in \figsref{fig:matching_problem}f--h show that $x_c$ ($\sim1151$) km is smaller than the prescribed $x_0$ (=1500 km).
Since the non-linear solution is only an approximation, the physical constraints are not exactly satisfied, with discontinuities around $x=x_c$, but the occurrence of the spurious negative precipitation region is minimized.

\begin{figure*}[h]
\centerline{\includegraphics[width=0.8\textwidth]{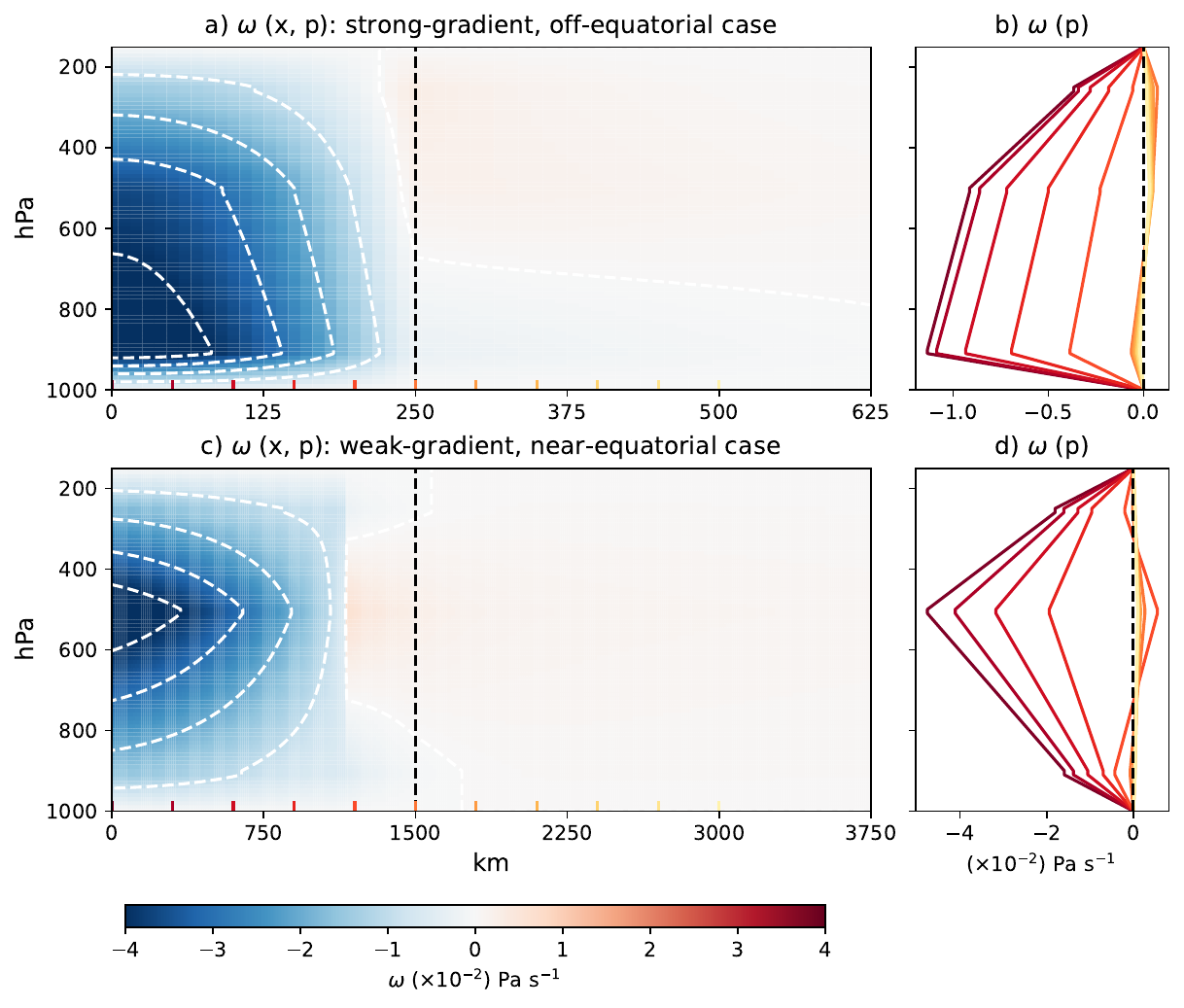}}
\caption{ a,b) contours and c,d) profiles of vertical velocity for the a,b) strong-gradient, off-equatorial and c,d) weak-gradient, near-equatorial cases. In panels a and b, the colored dashed lines on the x-axis mark values of $x$ for which the corresponding profiles of omega are shown in panels b and d. The black dashed line in a) and b) marks $x=x_c$. }\label{fig:omega_profiles}
\end{figure*}

For the weak-gradient, off-equatorial case, the tropospheric temperature perturbations (\figref{fig:matching_problem}f) do not resemble the $T_s$ forcing (\figref{fig:matching_problem}e) even in the boundary layer, suggesting that the free modes play a greater role in the full solution---unlike the strong-gradient case. The water vapor perturbations are nearly-equal in both the boundary layer and the lower-free troposphere (\figref{fig:matching_problem}g). This is consistent with the same sign convergence occurring in the boundary layer and lower-free troposphere (\figref{fig:omega_profiles}d). Compensating  upper-level divergence (\figref{fig:matching_problem}h) is larger relative to the strong-gradient case, and the resulting descent in the non-convective zone is sufficient to cause negative moisture perturbations in both boundary layer and lower-free troposphere.

The omega profiles as a function of distance from $x=0$ for strong- and weak-gradient cases are shown in \figref{fig:omega_profiles}. In both cases, strong ascent exists in the convective zone, which transitions to weak descent in the non-convective zone. However, the omega profile shapes show clear differences. In the presence of a strong $T_s$ gradient, the ascending omega profiles are bottom-heavy (\figsref{fig:omega_profiles}a and b). In contrast, the weak-gradient case displays more top-heavy profiles (\figsref{fig:omega_profiles}c and d). This dichotomy in omega profile shapes is reminiscent of the observed differences between the East and West Pacific Oceans \citep{back2006geographic}. The results from this simple model suggest that the shape of the surface temperature profile and the Coriolis force, to a large extent, control the top- versus bottom-heaviness of these vertical velocity profiles. Higher order vertical modes appear near the transition region between convective and non-convective zones. In both cases, a region of weak, low-level ascent extends into the non-convective zone before transitioning to deep descent. The region of low-level ascent adjacent to column-wide ascent is reminiscent of shallow and congestus clouds abutting deep convective clouds \citep{huaman2022assessing}. 

\section{Discussion}

\subsection{Baroclinic modes}

The first and second baroclinic omega profiles have long been long identified as the leading statistical modes---usually empirical orthogonal functions (EOFs)---of tropical omega profiles \citep{back2009relationship,hagos2010building, handlos2014estimating, hannah2016lagrangian, inoue2020vertical}. The ubiquitous first-baroclinic omega profile can also also be derived using the properties of a neutrally-buoyant, moist adiabatic plume \citep{neelin2000quasi, singh2022interaction, dang2024solutions}. The second-baroclinic mode profile is not derived from prior physical constraints, but imposed upon the problem \citep{mapes2000convective, khouider2006simple, kuang2008moisture}. 

In this study, the first- and second-baroclinic modes emerge emerge as the free modes of an $n$-layered model, thus providing physical reasons for why these modes should appear in data. Each baroclinic mode also has an associated horizontal scale, and higher-order vertical modes have much smaller spatial scales. This potentially explains why higher-order vertical modes are not observed in EOF analyses of omega profiles, since their contributions to omega variance expected to be small. The ubiquity of the first-baroclinic mode and its associated large horizontal scale is associated with the weak (but non-zero) free-tropospheric temperature gradients \citep[WTG][]{sobel2001weak} in the tropics. The tropospheric temperature maximizes near the region of strong convection and decays away from it, consistent with the picture of tropical `circus tents' \citep{williams2023circus}. As the forced problem in this work demonstrates, the boundary layer temperature gradients closely follow the surface temperature pattern, and thus may not not obey WTG, particularly in regions with strong surface temperature gradients.

% For idealized studies of tropical phenomena involving independent boundary layer and free-tropospheric dynamics, it might be cleaner to work with a multi-layered model than a multi-modal model. 

\subsection{Surface versus free-tropospheric control on convection}

A long standing debate in tropical meteorology centers on the relative roles for surface versus free-tropospheric control of tropical convection \citep{gill1980some,lindzen1987role, neelin1987modeling, chiang2001relative, back2009simple, sobel2006boundary, sobel2007simple, bunge2024variable}. Evidence for both mechanisms exists. The evidence for surface-driven convection hinges on the fact that in some regions, boundary-layer convergence can be diagnosed from SST gradients \citep{lindzen1987role, back2009simple, bunge2024variable}. Predominantly bottom-heavy convection is thought to co-occur with strong surface convergence. In other regions, free-tropospheric convection can contribute to the strength of the surface winds \citep{gill1980some, chiang2001relative}. This free-tropospheric controlled convection is thought to produce deeper omega profiles.

Part of the reason for this debate is the artificial separation of the troposphere into a boundary layer and a free troposphere, demarcated by a boundary layer top \citep[e.g.,][]{lindzen1987role, sobel2006boundary, back2009simple}. Impacts from either side of the boundary layer top are then used to argue for the influence of one layer or the other. However, it should be recognized that convection and radiation couple the tropospheric layers together. So a full answer to this debate must work with the entire troposphere. In the present model, the model is forced by an external surface temperature perturbation. So it is always `surface-driven'. However, the resulting solution can strongly project onto a more bottom- or top-heavy solution depending on the horizontal scale of the forcing, thus reproducing the strong surface convergence in one case, and strong free-tropospheric heating in another. In the steady (adjusted) state, it is therefore not meaningful to argue for the primacy of one layer over another. The present results suggest that it instead might be more meaningful to discuss the primacy of one mode over another. 

\section{Summary}

A simple model is introduced to understand the observed differences in omega profiles between the tropical East and West Pacific Oceans. We find the following model characteristics are sufficient: a steady-state model  with  linear equations---momentum, dry static energy and moisture---for $n$-layers.  Convective heating and drying are parameterized using the linear response functions from a cloud-resolving model \cite{kuang2010linear}, radiation with the linearized RRTMG model \citep{iacono2008radiative} and surface fluxes with linearized bulk formulas. A choice of $n=4$ layers is used to illustrate the results here and an f-plane facilitates analytic solutions. There exist $2n$ baroclinic eigenmodes in the $n$-layered model, and each mode is associated with a characteristic horizontal scale that yields decay away from a source. These scales  characterize the interplay between  stratification (which would yield wave propagation in inviscid time-dependent model),  the vertical communication of temperature by convection and radiation and the effect of surface forcing via fluxes.

The first and second baroclinic modes of this model have the two largest associated horizontal scales on the order of thousand(s) and hundreds of kilometers, respectively, and thus tend to dominate large-scale response. When the model is forced with a spatially-varying SST pattern, the character of the forced response thus shows dependence on the horizontal scale of the forcing, and to some extent, on the strength of the Coriolis force. Specifically, a forcing on the scale of hundreds of kilometers (strong horizontal gradient) excites a more bottom-heavy response in omega, while a forcing on the order of a thousand kilometers (weak horizontal gradient) excites a more top-heavy response. This also holds for a localized  SST anomaly with local convective and remote non-convective zones, where the deep mode tends to yield  greater far-field response relevant to teleconnections. 

Overall, the simple model introduced here appears capable of reproducing the salient inter-basin differences in tropical omega profiles.  The differences  between Eastern and Western Pacific are seen not as requiring different models or mechanisms. These differences emerge simply because the different characteristic scales of the leading vertical modes yield different responses to SST patterns.

% \clearpage
%%%%%%%%%%%%%%%%%%%%%%%%%%%%%%%%%%%%%%%%%%%%%%%%%%%%%%%%%%%%%%%%%%%%%
% ACKNOWLEDGMENTS
%%%%%%%%%%%%%%%%%%%%%%%%%%%%%%%%%%%%%%%%%%%%%%%%%%%%%%%%%%%%%%%%%%%%%
\acknowledgments
This work was supported in part by National Science Foundation grants AGS-2225956 and AGS-2414576. We thank Quentin Nicholas for providing the linear response functions. Conversations with Larissa Back helped improve this manuscript.
%%%%%%%%%%%%%%%%%%%%%%%%%%%%%%%%%%%%%%%%%%%%%%%%%%%%%%%%%%%%%%%%%%%%%
% DATA AVAILABILITY STATEMENT
%%%%%%%%%%%%%%%%%%%%%%%%%%%%%%%%%%%%%%%%%%%%%%%%%%%%%%%%%%%%%%%%%%%%%
% 
%
\datastatement
The scripts used to construct the model and produce the manuscript figures can be found at this \href{https://github.com/ahmedfiaz/linear\_layered\_model/tree/main}{GitHub repository}.

%%%%%%%%%%%%%%%%%%%%%%%%%%%%%%%%%%%%%%%%%%%%%%%%%%%%%%%%%%%%%%%%%%%%%
% REFERENCES
%%%%%%%%%%%%%%%%%%%%%%%%%%%%%%%%%%%%%%%%%%%%%%%%%%%%%%%%%%%%%%%%%%%%%
%  This shows how to enter the commands for making a bibliography using
%  BibTeX. It uses references.bib and the ametsocV6.bst file for the style.

\bibliographystyle{ametsocV6}
\bibliography{references}

\end{document}